\documentclass[a4paper,11pt]{article}
\usepackage{hyperref}
\usepackage{cite}
\usepackage{color}
\usepackage{amssymb}
\usepackage{amsfonts}
\usepackage{graphicx}
\usepackage{epstopdf}
\usepackage{dcolumn}
\usepackage{amsmath}
\usepackage{latexsym,bm}
\usepackage{geometry}

\def \be {\begin{equation}}
\def \ee {\end{equation}}
\def \bea {\begin{eqnarray}}
\def \eea {\end{eqnarray}}
\def \nn {\nonumber}

\def \a {\alpha}
\def \b {\beta}

\def \G {\Gamma}
\def \d {\delta}

\def \m {\mu}
\def \n {\nu}
\def \k {\kappa}

\def \s {\sigma}
\def \r {\rho}
\def \o {\omega}

\def \th {\theta}
\def \Th {\Theta}

\def \t {\tau}
\def \dag {\dagger}
\def \p {\partial}

\def\bd{\begin{document}}
\def\ed{\end{document}}
\def\nn{\nonumber}
\def\bea{\begin{eqnarray}}
\def\eea{\end{eqnarray}}
\let\bm=\bibitem
\let\la=\label

\def\N{{\cal N}}
\def\sst{\scriptscriptstyle}
\def\thetabar{\bar\theta}
\def\Tr{{\operatorname{Tr}}}
\def\one{\mbox{1 \kern-.59em {\rm l}}}

\def\a{\alpha}      \def\da{{\dot\alpha}}
\def\b{\beta}       \def\db{{\dot\beta}}
\def\c{\gamma}  \def\C{\Gamma}  \def\cdt{\dot\gamma}
\def\d{\delta}  \def\D{\Delta}  \def\ddt{\dot\delta}
\def\e{\epsilon}        \def\vare{\varepsilon}
\def\f{\phi}    \def\F{\Phi}    \def\vvf{\f}
\def\h{\eta}
\def\k{\kappa}
\def\l{\lambda} \def\L{\Lambda}
\def\m{\mu} \def\n{\nu}
\def\o{\omega}
\def\P{\Pi}
\def\r{\rho}
\def\s{\sigma}  \def\S{\Sigma}
\def\t{\tau}
\def\th{\theta} \def\Th{\Theta} \def\vth{\vartheta}
\def\X{\Xeta}
\def\z{\zeta}
\def\w{\wedge}
\def\u{\underline}
\def\hs{\hspace}

\def\cA{{\cal A}} \def\cB{{\cal B}} \def\cC{{\cal C}}
\def\cD{{\cal D}} \def\cE{{\cal E}} \def\cF{{\cal F}}
\def\cG{{\cal G}} \def\cH{{\cal H}} \def\cI{{\cal I}}
\def\cJ{{\cal J}} \def\cK{{\cal K}} \def\cL{{\cal L}}
\def\cM{{\cal M}} \def\cN{{\cal N}} \def\cO{{\cal O}}
\def\cP{{\cal P}} \def\cQ{{\cal Q}} \def\cR{{\cal R}}
\def\cS{{\cal S}} \def\cT{{\cal T}} \def\cU{{\cal U}}
\def\cV{{\cal V}} \def\cW{{\cal W}} \def\cX{{\cal X}}
\def\cY{{\cal Y}} \def\cZ{{\cal Z}}

\def\ua{\underline{\alpha}} \def\ubb{\underline{\beta}}
\def\ug{\underline{\gamma}}
\def\ub{\underline{\phantom{\alpha}}\!\!\!\beta}
\def\uc{\underline{\phantom{\alpha}}\!\!\!\gamma}
\def\um{\underline{\mu}} \def\un{\underline{\nu}}
\def\ud{\underline\delta}
\def\ue{\underline\epsilon}
\def\una{\underline a}\def\unA{\underline A}
\def\unb{\underline b}\def\unB{\underline B}
\def\unc{\underline c}\def\unC{\underline C}
\def\und{\underline d}\def\unD{\underline D}
\def\une{\underline e}\def\unE{\underline E}
\def\unf{\underline{\phantom{e}}\!\!\!\! f}\def\unF{\underline F}
\def\unm{\underline m}\def\unM{\underline M}
\def\unn{\underline n}\def\unN{\underline N}
\def\unp{\underline{\phantom{a}}\!\!\! p}\def\unP{\underline P}
\def\unq{\underline{\phantom{a}}\!\!\! q}
\def\unQ{\underline{\phantom{A}}\!\!\!\! Q}
\def\unH{\underline{H}}
\def\ul{\underline}

\def\As {{A \hspace{-6.4pt} \slash}\;}
\def\bs {{b \hspace{-6.4pt} \slash}\;}
\def\Ds {{D \hspace{-6.4pt} \slash}\;}
\def\ds {{\del \hspace{-6.4pt} \slash}\;}
\def\ss {{\s \hspace{-6.4pt} \slash}\;}
\def\ks {{ k \hspace{-6.4pt} \slash}\;}
\def\ps {{p \hspace{-6.4pt} \slash}\;}
\def\pas {{{p_1} \hspace{-6.4pt} \slash}\;}
\def\pbs {{{p_2} \hspace{-6.4pt} \slash}\;}

\def\Fh{\hat{F}}
\def\Vh{\hat{V}}
\def\Xh{\hat{X}}
\def\ah{\hat{a}}
\def\xh{\hat{x}}
\def\yh{\hat{y}}
\def\ph{\hat{p}}
\def\xih{\hat{\xi}}

\def\psit{\tilde{\psi}}
\def\Psit{\tilde{\Psi}}
\def\tht{\tilde{\th}}

\def\At{\tilde{A}}
\def\Qt{\tilde{Q}}
\def\Rt{\tilde{R}}
\def\Nt{\tilde{N}}

\def\at{\tilde{a}}
\def\st{\tilde{s}}
\def\ft{\tilde{f}}
\def\pt{\tilde{p}}
\def\qt{\tilde{q}}
\def\vt{\tilde{v}}
\def\nt{\tilde{n}}

\def\delb{\bar{\partial}}
\def\bz{\bar{z}}
\def\bD{\bar{D}}
\def\bB{\bar{B}}

\def\bk{{\bf k}}
\def\bl{{\bf l}}
\def\bp{{\bf p}}
\def\bq{{\bf q}}
\def\br{{\bf r}}
\def\bx{{\bf x}}
\def\by{{\bf y}}
\def\bR{{\bf R}}
\def\bV{{\bf V}}

\def\d{\delta}\def\D{\Delta}\def\ddt{\dot\delta}

\def\p{\partial} \def\del{\partial}
\def\xx{\times}
\def\uno{\mbox{1 \kern-.59em {\rm l}}}

\def\trp{^{\top}}
\def\inv{^{-1}}
\def\dag{{^{\dagger}}}

\def\pr{\prime}
\def\rar{\rightarrow}
\def\lar{\leftarrow}
\def\lrar{\leftrightarrow}

\title{\bf M5-branes in AdS$_4 \times Q^{1,1,1}$ spacetime}
\author{
De-Sheng Li$^{1}$\footnote{lidesheng@ihep.ac.cn}, Zheng-Wen Liu$^2$\footnote{zhengwen@ruc.edu.cn}, Jun-Bao Wu$^{1, 6}$\footnote{wujb@ihep.ac.cn}~~and Bin Chen$^{3, 4, 5, 6}$\footnote{bchen01@pku.edu.cn}
}
\date{}
\begin{document}
\maketitle
\begin{center}
{{\it
$^1$Institute of High Energy Physics,
and Theoretical Physics Center for Science Facilities,\\
Chinese Academy of Sciences, 19B Yuquan Road,
Beijing 100049, P.~R.~China\\
\vspace{2mm}
$^2$ Department of Physics, Renmin University of China, Beijing 100872, P.~R.~China\\
\vspace{2mm}
$^3$Department of Physics and State Key Laboratory of Nuclear Physics and Technology, Peking University, No. 5 Yiheyuan Rd, Beijing 100871, P.R.\! China\\
\vspace{2mm}
$^4$Center for High Energy Physics, Peking University, No. 5 Yiheyuan Rd, \\Beijing 100871, P.~R.~China\\
\vspace{2mm}
$^5$Beijing Center for Mathematics and Information Interdisciplinary Sciences,105 W~3rd~Ring Rd N, Beijing~100048, P.~R.~China\\
\vspace{2mm}
$^6$ Kavli Institute for Theoretical Physics China, CAS, Beijing 100190, P.~R.~China\\}
}
\vspace{10mm}
\end{center}

\date{}
\begin{abstract}

In this paper, we study the M5-brane configurations in AdS$_4 \times Q^{1,1,1}$ spacetime. We consider the configurations with an AdS$_2$ factor embedding into AdS$_4$, and manage to construct two solutions, which could be dual to line defects in the boundary gauge theory. Moreover we discuss their BPS nature and find that neither of them is supersymmetric. We show that the M5-brane with an R$_t$ or an AdS$_3$ factor found previously is half-BPS.
\end{abstract}
\thispagestyle{empty}

\newpage
\section{Introduction}

Great progress has been made on understanding the low energy effective action of $N$ M2-branes at the large $N$ limit since the construction of ABJM theory \cite{ABJM},
which was in part inspired by \cite{BL1,BL2,BL3,G1,G2} among other works. ABJM theory is a three-dimensional ${\cal N}=6$ super-Chern-Simons theory with gauge group
 $U(N)_k\times U(N)_{-k}$. This theory is dual to M-theory on $AdS_4\times S^7/Z_k$ or type IIA theory on $AdS_4\times \mathbb{CP}^3$. By solving the ABJM matrix model obtained via
supersymmetric localization \cite{Kapustin:2009kz},
we finally had a satisfying understanding \cite{Drukker:2010nc} of the scaling behavior  $N^{3/2}$  for the counting of the degrees of freedom for
$N$ M2-branes, first obtained through the computations on the gravity side \cite{Klebanov:1996un}.  
Many examples of $AdS_4/CFT_3$ correspondence with less supersymmetries have been studied as well. On the gravity side, the correspondence involves M-theory on $AdS_4\times Y^7$, with $Y^7$ being certain manifolds or orbifolds. 
The dual field theory can be  three-dimensional Chern-Simons-matter theory with ${\cal N}=1$ ($2, 3$) supersymmetries when $Y^7$ is a weak $G_2$ (Sasaki-Einstein, $3$-Sasaki) manifold (or its orbifold, preserving the same amount of supersymmetries)  \cite{Acharya:1998db, Morrison:1998cs}. The three-dimensional Chern-Simons-matter field theories with ${\cal N}=4, 5, 6$ supersymmetries, corresponding to certain orbifolds $Y^7$, have also been studied in \cite{ABJM,Hosomichi:2008jb, ABJ, Imamura:2008nn, Terashima:2008ba, Benna:2008zy}. Among them, the study of M-theory on $AdS_4\times Q^{1, 1, 1}$ \cite{Fabbri:1999hw,Franco:2008um,Franco:2009sp,Aganagic:2009zk,Benini,Jafferis} is of particular interest, because the metric of Sasaki-Einstein manifold $Q^{1, 1, 1}$ is quite simple. 

In M-theory, there are two kinds of nonperturbative objects: M2-brane and M5-brane. Their roles in AdS$_4$/CFT$_3$ correspondence are not completely clear. For example, the dimension reduction of M2-brane to ten dimensions may give us a fundamental string, which could be dual to the Wilson loop in the field theory\cite{Rey, Mal98}. However, though the simplest embedding of F-strings inside the dual IIA string theory background $AdS_4\times \mathbb{CP}^3$ is half BPS \cite{Drukker, Rey2}, the field theory construction of the BPS Wilson loop operator is highly nontrivial\cite{Drukker, Rey2, GaiottoYin, Berenstein:2008dc, Chen, drukker2, LeeLee}. Less supersymmetric Wilson loops in ABJM theory were studied in \cite{Griguolo:2012iq, Cardinali:2012ru, Kim:2013oza, Correa:2014aga}. General studies on Wilson loops in ${\cal N}=2$ super-Chern-Simons theory were performed in \cite{Sparks}. Very recently, the  BPS M2-branes in $AdS_4\times Q^{1,1,1}$ dual to BPS Wilson loops and vortex loops were studied in \cite{WuZhu} based on explicit expressions of Killing spinors. Other types of membranes in $AdS_4\times Q^{1, 1, 1}$ were studied in  \cite{Klebanov:2010tj, Benishti:2010jn, Kim}.

Besides M2-branes, there are also M5-branes in M-theory. In the context of AdS/CFT correspondence, M5-brane could be dual to the baryonic operator or the defects, including the line defect and the domain wall in the field theory. It may also appear due to the Myers' polarization effect of multiple M2-branes\cite{Myers:1999,Chen:2007ir,Lunin:2007ab}. It is not easy to find the M5-brane configuration in curved spacetime because its equations of motion are hard to solve. In the case of $AdS_4\times Q^{1, 1, 1}$, the M5-branes with an ${\bf R}_t$ and $AdS_3$ factor have been studied  in \cite{Ahn:1999ec, Benishti:2010jn}. The first M5-brane is dual to a certain baryonic operator, while the second one is dual to the domain wall. An unanswered question on these solutions is whether or not they are supersymmetric. In \cite{Fujita:2009kw}, a certain $D4$-brane in $AdS_4\times  \mathbb{CP}^3$ was found as the holographic dual of the supersymmetric domain wall in ABJM theory.

The main topics of this paper is to study M5-brane solutions and their BPS nature in $AdS_4\times Q^{1, 1, 1}$ spacetime. We pay special attention to  M5-branes whose worldvolume includes an $AdS_2$ factor. These M5-branes should be dual to certain one-dimensional defects in the dual gauge theory, though they may not be dual to the Wilson loop operator or the vortex loop operator \cite{drukker4}. With the projection condition on the Killing spinor in mind, we make two kinds of ansatz which have the potential to be supersymmetric. However, after solving the M5-brane equations of motion and studying the BPS conditions for M5-branes, we find that none of them is BPS. This shows that it is quite hard to find BPS M5-branes with an $AdS_2$ factor in $AdS_4\times Q^{1,1,1}$. Besides, we check the supersymmetries preserved by the previously mentioned M5-brane with an ${\bf R}_t$ or $AdS_3$ factor and
find that both of them are half-BPS.

In the next section, we will briefly review M5-brane equations of motion and the projection condition for the supersymmetries preserved by the
probe M5-brane. In section 3, we introduce the background fields and the Killing spinors of M-theory on $AdS_4\times Q^{1, 1, 1}$. In section 4, we present
two M5-brane solutions whose worldvolumes involve an $AdS_2$ factor.  In section 5, we discuss the supersymmetries preserved by M5-brane with an ${\bf R}_t$ or $AdS_3$ factor.
We conclude this paper with some brief discussions. In the appendix, we gather the explicit form of the connection coefficients used in the main text.

\section{M5-brane equations of motion}

Various proposals and aspects of M5-brane actions have been studied in \cite{Sezgin97, Sezgin99, Howe96, Chu97, Sundell97, Sorokin97} (for a
review of M-theory branes, see \cite{Berman}). In this section we briefly review the covariant equations of motion for M5-branes \cite{Sezgin97} and
the supersymmetric conditions for the probe M5-brane.

The massless bosonic fields of $11$-dimensional M-theory include the metric\footnote{Our notation is as follows: indices from the
beginning(middle) of the alphabet refer to frame(coordinate)
indices, and the underlined indices refer to target space ones.}
\be ds^2_{11}=g_{\ul{m}\ul{n}}dx^{\ul{m}}dx^{\ul{n}},\ee
and the $4$-form field strength \be H_4=H_{\ul{m_1}\cdots \ul{m_4}}dx^{\ul{m_1}}\wedge \cdots \wedge dx^{\ul{m_4}}.\ee
We also need the target space vielbein $E^{\ul{a}}_{\ul{m}}$ satisfying
\be E^{\ul{a}}_{\ul{m}}E^{\ul{b}}_{\ul{n}}\eta_{\ul{a}\ul{b}}=g_{\ul{m}\ul{n}},\ee
and the Hodge dual of $H_4$ denoted as $H_7$ whose components are $H_{\ul{m_1}\cdots \ul{m_7}}$.

The probe M5-brane solution is described in terms of the embedding $x^{\ul{m}}(\xi^m)$ and a  self-dual $3$-form field $h_{mnp}$
on the M5-brane worldvolume. Here $\xi^m, m=0, \ldots, 5$ are coordinates of the worldvolume. From the embedding, we can define
the induced metric \be g_{mn}={\cal E}^{\ul{a}}_m{\cal E}^{\ul{b}}_n\eta_{\ul{a}\ul{b}}, \ee
with \be{\cal E}^{\ul{a}}_m=\partial_mx^{\ul{n}}E^{\ul{a}}_{\ul{n}}.\ee
Starting with $h_{mnp}$, which is self-dual with respect to this induced metric, we define the following list of quantities
\bea
 k_m^{~n}&=&h_{mpq}h^{npq}, \\
 Q&=&1-\frac{2}{3}k_m^{~n}k_n^{~m}, \\
 m_p^{~q}&=&\delta_p^{q}-2k_p^{~q}, \\
 H_{mnp}&=&4Q^{-1}(1+2k)_m^{~q}h_{qnp},\\
 G^{mn}&=&\Big(1+\frac{2}{3}k^2\Big)g^{mn}-4k^{mn},\\
 P_{\underline a}^{~\underline c}&=&\delta^{\underline
 c}_{\underline a}-\cE_{\underline a}^m\cE_m^{~{\underline c}},\\
 Y_{mn}&=&\big(4\star {\underline H}-2(m\star {\underline H}+\star {\underline H}m)+m\star {\underline H}m\big)_{mn},
 \eea
where \be \star {\underline
H}^{mn}=\frac{1}{4!\sqrt{-g}}\epsilon^{mnpqrs}{\underline
H}_{pqrs}. \ee
The covariant derivative $\nabla_m{\cal E}^{\ul{c}}_n$ is defined as
\be \nabla_m{\cal E}^{\ul{c}}_n=\partial_m{\cal E}^{\ul{c}}_n-{\Gamma_{~mn}^p}{\cal E}^{\ul{c}}_p+{\cal E}^{\ul{a}}_m{\cal E}^{\ul{b}}_n
{\omega^{\ul{c}}_{~\ul{a}\ul{b}}},\ee
where ${\G^p_{~mn}}$ is the Christoffel symbol with respect to the induced metric on the worldvolume  and ${\o^{\underline c}_{~{\underline a}{\underline
 b}}}$ is the spin connection of the background spacetime.

After defining these quantities, the  equations of motion of an M5-brane include three parts:
\begin{itemize}
\item Bianchi identity
\be dH_3=-P[H_4],\ee
where $P[H_4]$ is the pull-back of the target space 4-form flux.
\item{Scalar equation}
\be
 G^{mn}\nabla_m \cE_n^{\underline
 c}=\frac{Q}{\sqrt{-g}}\epsilon^{m_1\cdots
 m_6}\bigg(\frac{1}{6!}H^{\underline a}_{~m_1\cdots
 m_6}+\frac{1}{(3!)^2}H^{\underline
 a}_{~m_1m_2m_3}H_{m_4m_5m_6}\bigg)P_{\underline
 a}^{~\underline c},
 \ee
\item{Tensor equation}
\be G^{mn}\nabla_mH_{npq}=Q^{-1}\big(4Y-2(mY+Ym)+mYm\big)_{pq}.
 \ee
\end{itemize}

To study the supersymmetries preserved by the probe M5-brane, we need to solve the kappa
symmetry projection condition
\be \Gamma_{M5}\eta=\eta,\ee
where $\eta$ is the solution of the Killing spinor equation of the M-theory background
\be \nabla_{\ul{m}}\eta+\frac1{576}\big(3\Gamma_{{\ul n}{\ul p}{\ul q}{\ul r}}\Gamma_{\ul m}-\Gamma_{\ul m}\Gamma_{{\ul n}{\ul p}{\ul q}{\ul r}}\big)H^{{\ul n}{\ul p}{\ul q}{\ul r}}\eta=0, \ee
and $\Gamma_{M5}$ is determined by the
embedding of M5-brane and the flux on it \cite{Sezgin97}
 \begin{equation}
 \Gamma_{M5}=\frac{1}{6!\sqrt{-g}}\epsilon^{j_1\cdots j_6}\Big(\Gamma_{<j_1\cdots
 j_6>}+40\Gamma_{<j_1j_2j_3>}h_{j_4j_5j_6}\Big).
 \end{equation}
Here $g$ is the determinant of the induced worldvolume metric
component, and $h_{j_4j_5j_6}$ is the self-dual 3-form on the
M5-brane. $\Gamma_{<j_1\cdots j_n>}$ is defined as
 \begin{equation}
 \Gamma_{<j_1 \cdots j_n>}={\cal E}^{\underline a_1}_{j_1}\cdots{\cal E}^{\underline
 a_n}_{j_n}\Gamma_{{\underline a_1}\cdots {\underline a_n}},
 \end{equation}
 where $\Gamma_{{\underline a_1}\cdots {\underline a_n}}$ is the antisymmetrized product of the Gamma matrices
 in the orthonormal frame.

\section{Background fields and Killing spinors}
The metric on $AdS_4\times Q^{1, 1, 1}$ is
\bea ds^2&=&R^2(ds^2_4+ds^2_7),\\
ds^2_4&=&\frac14\big(\cosh^2u(-\cosh^2\rho dt^2+d\rho^2)+du^2+\sinh^2u d\phi^2\big),\\
ds^2_7&=&\frac18\sum_{i=1}^3\big(d\theta^2_i+\sin^2\theta_id\phi_i^2\big)+\frac1{16}\bigg(d\psi+\sum_{i=1}^3\cos\theta_i d\phi_i\bigg)^2,\eea
with $\theta_i\in[0, \pi], \phi_i\in [0, 2\pi] \, (i=1, 2, 3), \psi\in [0, 4\pi]$.
The four-form field strength on this background is
\be H_4=\frac{3R^3}8 \cosh^2u\sinh u\cosh\rho dt\w d\rho \w du \w d\phi. \ee
The vielbeins of the eleven-dimensional metric are
\begin{align}\label{vielbein}
\begin{aligned}
  e^{\ul{0}} \,&=\, \frac{R}2\cosh u\cosh\rho dt, \\
  e^{\ul{2}} \,&=\, \frac{R}2 du, \\
  e^{\ul{4}} \,&=\, \frac{R}{2\sqrt{2}} d\theta_1, \\
  e^{\ul{6}} \,&=\, \frac{R}{2\sqrt{2}} d\theta_2, \\
  e^{\ul{8}} \,&=\, \frac{R}{2\sqrt{2}}d\theta_3, \\
  e^{\ul{\sharp}} \,&=\, \frac{R}4 \Big(d\psi+\sum_{i=1}^3\cos\theta_id\phi_i\Big),
\end{aligned}\quad
\begin{aligned}
  e^{\ul{1}} \,&=\, \frac{R}2\cosh u d\rho, \\
  e^{\ul{3}} \,&=\, \frac{R}2 \sinh u d\phi, \\
  e^{\ul{5}} \,&=\, \frac{R}{2\sqrt{2}} \sin\theta_1d\phi_1, \\
  e^{\ul{7}} \,&=\, \frac{R}{2\sqrt{2}} \sin\theta_2d\phi_2, \\
  e^{\ul{9}} \,&=\, \frac{R}{2\sqrt{2}} \sin\theta_3d\phi_3, \\
  \\ \\
\end{aligned}
\end{align}
such that $H_4$ can now be written as
\be H_4=\frac{6}{R} e^{\ul{0}}\w e^{\ul {1}} \w e^{\ul{2}}\w e^{\ul{3}}. \ee
As mentioned in the previous section, the Killing spinors of $AdS_4\times Q^{1, 1, 1}$ satisfy the following equation:
\be \nabla_{\ul{m}}\eta+\frac1{576}\big(3\Gamma_{{\ul n}{\ul p}{\ul q}{\ul r}}\Gamma_{\ul m}-\Gamma_{\ul m}\Gamma_{{\ul n}{\ul p}{\ul q}{\ul r}}\big)H^{{\ul n}{\ul p}{\ul q}{\ul r}}\eta=0. \ee
Our convention about the product of the eleven $\Gamma$ matrices is \be\Gamma_{\ul{0}\ul{1}\ul{2}\ul{3}\ul{4}\ul{5}\ul{6}\ul{7}\ul{8}\ul{9}\ul{\sharp}}=1.\ee
Using the vielbeins  given above and the spin connections given in the Appendix, we find that the solution to the above equation is\footnote{The Killing spinor in a slightly different moving frame was given in \cite{WuZhu}.The Killing spinors of $Q^{1, 1, 1}$ were also studied previously in \cite{Hoxha, Donos}. The Killing spinors of $AdS_4$ were given in this coordinate system in \cite{Drukker}.}
\be \eta=e^{\frac{u}2\Gamma_{\ul{2}}\hat{\Gamma}}e^{\frac{\rho}2\Gamma_{\ul{1}}\hat{\Gamma}}e^{\frac{t}2\Gamma_{\ul{0}}\hat{\Gamma}}e^{\frac{\phi}2\Gamma_{{\ul{2}}{\ul{3}}}}
e^{-\frac\psi2\G_{\ul{45}}}\eta_0,\label{spinor} \ee
where $\eta_0$ is independent of all the coordinates 
and satisfies the projection conditions
\be\Gamma^{\ul{4}\ul{5}}\eta_0=\Gamma^{\ul{6}\ul{7}}\eta_0=\Gamma^{\ul{8}\ul{9}}\eta_0,\label{projector1}\ee
and $\hat{\Gamma}$ is defined as
\be \hat{\Gamma} \equiv \Gamma_{\ul{0}\ul{1}\ul{2}\ul{3}}. \ee

We will also need the metric of $AdS_4$ in the Poincar\'e coordinates
\be ds^2_4=\frac14\bigg(\frac{-dt^2+dx_1^2+dx_2^2+dy^2}{y^2}\bigg). \ee
Now the vielbeins in the $AdS_4$ part are
\begin{align}\label{}
\begin{aligned}
  e^{\ul{0}}\,&=\, \frac{R}2\frac{dt}{y},  \\
  e^{\ul{2}}\,&=\, \frac{R}2 \frac{dx_2}{y},
\end{aligned}\quad
\begin{aligned}
  e^{\ul{1}} \,&=\, \frac{R}2\frac{dx_1}{y}, \\
  e^{\ul{3}} \,&=\, \frac{R}2 \frac{dy}{y},
\end{aligned}
\end{align}
and the corresponding spin connections are
\be
 \omega^{\ul{0}\ul{3}}=-\frac2Re^{\ul{0}},\hs{3ex}
 \omega^{\ul{1}\ul{3}}=-\frac2Re^{\ul{1}},\hs{3ex}
 \omega^{\ul{2}\ul{3}}=-\frac2Re^{\ul{2}}.
\ee

In Poincar\'e coordinates, the solutions to the Killing spinor equations are
\begin{eqnarray} \eta=y^{1/2}\eta_++y^{-1/2}(\eta_-+x^{\ul{\mu}}\Gamma_{\ul{\mu3}}\eta_+). \label{ks2}\end{eqnarray}
Here $\eta_\pm=\exp\big(-\frac{\psi}{2}\Gamma_{\ul{45}}\big)\eta_{\pm}^0$, and $\eta_{\pm}^0$ satisfies \be\Gamma_{\ul{3}}\hat\Gamma\eta_\pm^0=\pm\eta_\pm^0, \quad\Gamma^{\ul{4}\ul{5}}\eta_\pm^0=\Gamma^{\ul{6}\ul{7}}\eta_\pm^0=\Gamma^{\ul{8}\ul{9}}\eta_\pm^0.\label{projector2}\ee

\section{M5-branes dual to line defects}

In this section we find two M5-brane solutions dual to line defects in the boundary gauge theory.
The first solution has an $AdS_2$ factor in the $AdS_4$ part of the background geometry, while the second
solution has an $AdS_2\times S^1$ factor in the $AdS_4$ part.

\subsection{The first solution}
For this solution, the topology of the worldvolume of the M5-brane is $AdS_2\times M^4$ with $AdS_2\subset AdS_4$ and $M^4\subset Q^{1, 1, 1}$.
The embedding of this M5-brane is
\begin{align}\label{}
  \xi^0 &= t,\quad \xi^1=\rho,  \\
  \xi^2=\theta_1,\quad \xi^3 &= \phi_1,\quad \xi^4=\theta_2,\quad \xi^5=\phi_2,
\end{align}
with other coordinates fixed.\footnote{Note, in particular, that $u$ takes a fixed value $u_0$.}
We choose the $3$-form $h_3$ to be zero.

Now the induced metric is
\begin{align}\label{}
   d\tilde{s}^2 \,=\, R^2\Bigg(&\frac14\cosh^2u_0(-\cosh^2\rho dt^2+d\rho^2)+\frac18\sum_{i=1}^2(d\theta_i^2+\sin^2\theta_id\phi_i^2)\nn\\
   &+\frac1{16}\bigg(\sum_{i=1}^2\cos\theta_id\phi_i\bigg)^2\Bigg).\label{induced}
\end{align}
The nonzero components of $\cE^{\ul{a}}_m$ are
\begin{align}\label{}
  \cE^{\ul{0}}_t &= \frac{R}2\cosh u_0\cosh\rho,\quad \cE^{\ul{1}}_\rho = \frac{R}2\cosh u_0, \\
  \cE^{\ul{4}}_{\theta_1} &= \cE^{\ul{6}}_{\theta_2} = \frac{R}{\sqrt{8}},\quad
  \cE^{\ul{5}}_{\phi_1} = \frac{R}{\sqrt{8}}\sin\theta_1,\quad
  \cE^{\ul{7}}_{\phi_2} = \frac{R}{\sqrt{8}}\sin\theta_2, \\
  \cE^{\ul{\sharp}}_{\phi_1} &= \frac{R}{4}\cos\theta_1,\quad
  \cE^{\ul{\sharp}}_{\phi_2} = \frac{R}{4}\cos\theta_2.
\end{align}

From $h_3=0$, it is easy to obtain that $H_3=0$ and $G_{mn}=g_{mn}$. Then the Bianchi identity and the tensor equations are satisfied trivially.
And after some computations, we find that the scalar equations give the constraint that $u_0=0$.

The Killing spinor on the worldvolume of this M5-brane is
\be\eta=e^{\frac\rho2 \Gamma_{\ul{1}}\hat{\Gamma}}e^{\frac{t}2\Gamma_{\ul{0}}\hat{\Gamma}}e^{\frac{\phi_0}2\Gamma_{\ul{23}}}e^{-\frac{\psi_0}2\Gamma_{\ul{45}}}\eta_0.\ee
In this case $\Gamma_{M5}$ is
\bea
\Gamma_{M5} =
\frac{\sqrt{2}\sin\theta_1\sin\theta_2\Gamma_{\ul{014567}}+\sin\theta_1\cos\theta_2\Gamma_{\ul{01456\sharp}}+\cos\theta_1\sin\theta_2\Gamma_{\ul{01467\sharp}}}
{\sqrt{\sin^2\theta_1+\sin^2\theta_2}}.
\eea
Considering the points in the submanifold $t=\rho=\theta_2=0, \theta_1=\pi/2$ on the worldvolume, $\Gamma_{M5}\eta=\eta$ becomes
\be \Gamma_{\ul{01456\sharp}}\eta_0=\eta_0. \ee
However, this projection condition is not compatible with the projection condition $\Gamma_{\ul{4567}}\eta_0 = -\eta_0$ from the Killing spinor equations.
This leads to the conclusion that this M5-brane is not supersymmetric.

\subsection{The second solution\label{2nd solution}}
The topology of the worldvolume of this M5-brane is $AdS_2\times S^1 \times M_3$ with $AdS_2\times S^1 \subset AdS_4$ and $M_3\subset Q^{1, 1, 1}$.
The embedding is
\bea \xi^0=t,\quad \xi^1=\rho,\quad \xi^2=\phi, \\
\xi^3=\theta_1,\quad \xi^4=\phi_1,\quad \xi^5=\psi,
\eea
with other coordinates fixed.
The $3$-form field $h_3$ is chosen to be
\be h_3=a(\xi)\bigg(\frac{R^3}{8}\cosh^2 u_0\sinh u_0\cosh\rho dt\wedge d\rho\wedge d\phi-\frac{R^3}{32}\sin\theta_1d\theta_1\wedge d\phi_1\wedge d\psi \bigg), \ee
satisfying the condition that $h_3=\ast h_3$.

The induced metric is
\begin{align}\label{}
d\tilde{s}^2 \,=\, R^2\bigg(&\frac14 \cosh^2u_0 (-\cosh^2\rho dt^2+d\rho^2)+\frac14\sinh^2u_0d\phi^2\nn\\
&+~\frac18 (d\theta_1^2+\sin^2\theta_1d\phi_1^2)+\frac1{16}(d\psi+\cos\theta_1d\phi_1)^2\bigg).\label{induced2}
\end{align}
The nonzero components of $\cE^{\ul{a}}_m$ are
\bea
\cE^{\ul{0}}_t=\frac{R}2\cosh u_0\cosh\rho, \quad \cE^{\ul{1}}_\rho=\frac{R}2\cosh u_0, \quad \cE^{\ul{3}}_\phi=\frac{R}2\sinh u_0,\\ \cE^{\ul{4}}_{\theta_1}=\frac{R}{\sqrt{8}}, \quad \cE^{\ul{5}}_{\phi_1}=\frac{R}{\sqrt{8}}\sin\theta_1, \quad
\cE^{\ul{\sharp}}_{\phi_1}=\frac{R}{4}\cos\theta_1,\quad \cE^{\ul{\sharp}}_{\psi}=\frac{R}{4}.
\eea

We list some important quantities for this solution here:
\be {k^{~m}_{n}}=\left(\begin{array}{cc}-2a^2 I_{3\times 3}&0\\
0&2a^2I_{3\times 3}\end{array}\right), \ee
\be Q=1-\frac23\Tr k^2=1-16a^4.\ee
The nonzero components of $G_{mn}$ are
\be  G_{mn}=
(1+4a^2)^2g_{mn},\ee
when $m, n\in \{t, \rho, \phi\}$, and
\be G_{mn}=(1-4a^2)^2g_{mn},\ee
when $m, n\in \{\theta_1, \phi_1, \psi\}$. And
\bea H_3 \,=\, \frac{aR^3 \cosh^2u_0\sinh u_0 \cosh\rho}{2(1+4a^2)} dt\wedge d\rho \wedge d\phi
 \,-\, \frac{aR^3 \sin\theta_1}{8(1-4a^2)} d\theta_1\wedge d\phi_1\wedge d\psi.\eea

The Bianchi identity gives
\be dH_3=0 \ee
which leads to the fact that $a$ is a constant. The tensor equations are automatically satisfied under
this condition.
By some computations, we find that scalar equations give the following relation between $a$ and $u_0$:
\be2\tanh u_0+\coth u_0=\frac{12a}{1+4 a^2}.\ee

On the worldvolume of this M5-brane, the Killing spinor reads
\be\eta=e^{\frac{u_0}2 \Gamma_{\ul{2}}\hat{\Gamma}} e^{\frac{t}2\Gamma_{\ul{0}}\hat{\Gamma}}e^{\frac{\phi}2\Gamma_{\ul{23}}}e^{-\frac{\psi}2\Gamma_{\ul{45}}}\eta_0.\ee
And $\Gamma_{M5}$ now becomes
\be\Gamma_{M5}=\Gamma_{\ul{01345\sharp}}-2a(\Gamma_{\ul{013}}+\Gamma_{\ul{45\sharp}}).\ee
By studying the special cases with $\rho=t=\phi=\psi=0$ and $\rho=t=\psi=0$, $\phi=\pi/2$, we find  that this M5-brane
is non-BPS.

\section{Supersymmetric M5-branes}
The M5-branes with an $AdS_2$ factor that we found in the last section are not supersymmetric. In this section, we discuss the BPS nature of two other M5-brane configurations proposed in the literature and find that they each keep half of the supersymmetries.

\subsection{M5-brane with an ${\bf R}_t$ factor}
The M5-brane with an ${\bf R}_t$ factor in $AdS_4$ and with five other directions in $Q^{1, 1, 1}$ was studied in \cite{Ahn:1999ec, Benishti:2010jn}.
Now we show explicitly that this brane configuration satisfies M5-brane equations of motion and preserves half of the supersymmetries
 of the $AdS_4\times Q^{1, 1, 1}$ background.
The embedding of this M5-brane is
\be\xi^0=t,\quad \xi^1=\theta_1,\quad \xi^2=\phi_1,\quad \xi^3=\theta_2,\quad \xi^4=\phi_2,\quad \xi^5=\psi,\ee
with $\theta_3, \phi_3, u, \rho, \phi$ fixed on the worldvolume, and the $3$-form field $h_3$ is chosen to be zero.
The induced metric on the worldvolume is
\bea d\tilde{s}^2&=& R^2\Bigg(-\frac{1}{4}\cosh^2u\cosh^2\rho dt^2+\frac{1}{8}\sum_{i=1}^2\big(d\theta_i^2+\sin^2\theta_id\phi_i^2\big)\nonumber\\
&&+~\frac{1}{16}\bigg(d\psi+\sum_{i=1}^2\cos\theta_id\phi_i\bigg)^2
\Bigg). \label{induced3}\eea

The nonzero components of $\cE^{\ul{a}}_m$ are
\bea \cE^{\ul{0}}_t = \frac{R}2\cosh u\cosh\rho, \\
\cE^{\ul{4}}_{\theta_1} = \cE^{\ul{6}}_{\theta_2} = \frac{R}{\sqrt{8}}, \quad
\cE^{\ul{5}}_{\phi_1}=\frac{R}{\sqrt{8}}\sin\theta_1, \quad
\cE^{\ul{7}}_{\phi_2}=\frac{R}{\sqrt{8}}\sin\theta_2, \\
\cE^{\ul{\sharp}}_{\phi_1}=\frac{R}{4}\cos\theta_1, \quad \cE^{\ul{\sharp}}_{\phi_2}=\frac{R}{4}\cos\theta_2, \quad
\cE^{\ul{\sharp}}_\psi=\frac{R}{4}.\eea

After some short computations, we obtain
\bea Q=1,\quad H_3=0,\quad G_{mn}=g_{mn}.\eea
Now the Bianchi identity and the tensor equation are satisfied automatically, and the scalar equations give the constraint that
\be u=\rho=0. \ee

Now we can easily obtain
\be \G_{M5}=\G_{\ul{04567\sharp}}. \ee
Using eqs.~(\ref{spinor}) and (\ref{projector1}), we find that the supersymmetric condition \be\G_{M5}\eta=\eta,\ee
is equivalent to the projection condition
\be \G_{\ul{0\sharp}}\eta_0=-\eta_0.\ee
Since this condition is compatible with the projection conditions in Eq.~(\ref{projector1}), we arrive at the conclusion that this M5-brane is half-BPS.

\subsection{M5-brane with an $AdS_3$ factor}

The M5-brane with an $AdS_3$ factor was studied in \cite{Ahn:1999ec} and was argued there to be dual to a domain wall in the field theory.
Here we show explicitly that this configuration does satisfy the equations of motion for probe M5-brane and, moreover, is half-BPS. We will also make contact
with general discussions on BPS M5-branes in $AdS_4\times Y^7$ background in \cite{Yamaguchi}.

Now we use the Poincar\'e coordinates of $AdS_4$. The embedding {of} this M5-brane is
\begin{align}\label{}
  \xi^0 &= t,\quad \xi^1=x_1,\quad \xi^2=y,\quad x_2=f(y), \\
  \xi^3 &= \theta_1,\quad \xi^4=\phi_1,\quad \xi^5=\psi.
\end{align}
The $3$-form field $h_3$ is chosen to be
\be h_3=a(\xi)\bigg(\frac{R^3 \sqrt{1+f^{\prime 2}}}{8}dy\wedge dt\wedge dx_1
-\frac{R^3 \sin\theta_1}{32}d\theta_1\wedge d\phi_1\wedge d\psi \bigg), \ee
satisfying the condition that $h_3=\ast h_3$.
The topology of the worldvolume of this M5-brane is $AdS_3\times M_3$. Notice that the $M_3$ part is the same as the one in subsection~\ref{2nd solution}.

The induced metric reads
\be ds^2=\frac{R^2}{4y^2}\big(-dt^2+dx_1^2+(1+f^{\prime 2})dy^2\big)
+\frac{R^2}{8}\big(d\theta_1^2+\sin^2\theta_1d\phi_1^2\big)+\frac{R^2}{16}\big(d\psi+\cos\theta_1d\phi_1\big)^2. \label{induced4}\ee
The nonzero components of $\cE^{\ul{a}}_m$ are
\bea \cE^{\ul{0}}_t=\frac{R}{2y},   \quad \cE^{\ul{1}}_{x_1}=\frac{R}{2y},  \quad \cE^{\ul{2}}_y=\frac{Rf^\prime}{2y},\quad \cE^{\ul{3}}_y=\frac{R}{2y},\\
 \cE^{\ul{4}}_{\theta_1} = \frac{R}{2\sqrt{2}},\quad \cE^{\ul{5}}_{\phi_1}=\frac{R\sin\theta_1}{2\sqrt{2}}, \quad\cE^{\ul{\sharp}}_{\phi_1}=\frac{R\cos\theta_1}{4},\quad \cE^{\ul{\sharp}}_\psi=\frac{R}4.\eea
Now we have
\be {k^{~m}_{n}}=\left(\begin{array}{cc}-2a^2 I_{3\times 3}&0\\
0&2a^2I_{3\times 3}\end{array}\right), \ee
\be Q=1-\frac23\Tr k^2=1-16a^4.\ee
The nonzero components of $G_{mn}$ are
\be  G_{mn}=
(1+4a^2)^2g_{mn},\ee
when $m, n\in \{t, x_1, y\}$, and
\be G_{mn}=(1-4a^2)^2g_{mn},\ee
when $m, n\in \{\theta_1, \phi_1, \psi\}$.
The nonzero components of {$P_{\ul{a}}^{~\ul{c}}$ } are
{\bea P_{\ul{2}}^{~\ul{2}}&=&\frac{1}{1+f^{\prime2}},\\
P_{\ul{3}}^{~\ul{3}}&=&\frac{f^{\prime2}}{1+f^{\prime2}},\\
P_{\ul{2}}^{~\ul{3}}&=&P_{\ul{3}}^{~\ul{2}}~=~-\frac{f^\prime}{1+f^{\prime2}}, \\
P_{\ul{6}}^{~\ul{6}}&=&P_{\ul{7}}^{~\ul{7}}~=~P_{\ul{8}}^{~\ul{8}}~=~P_{\ul{9}}^{~\ul{9}}~=~1.\eea  }

The $3$-form field $H_3$ is
\be H_3=\frac{aR^3\sqrt{1+f^{\prime 2}}}{2y^3(1+4a^2)} dy\wedge dt \wedge dx_1
- \frac{aR^3\sin\theta_1}{8(1-4a^2)} d\theta_1\wedge d\phi_1\wedge d\psi. \ee
The Bianchi identity implies that  $a$ should be a constant.
Under this condition, the tensor equations are satisfied and the only non-trivial condition given by the scalar equations is
\bea\frac{y}{\sqrt{1+f^{\prime2}}}\left(-\frac{3f^\prime}{y}+\frac{f^{\prime\prime}}{1+f^{\prime2}}\right)=\frac{12a}{1+4a^2}.\eea

For the special case, $f(y)=\kappa y$ with $\kappa$ a constant, we get
\be \frac{-\kappa}{\sqrt{1+\kappa^2}}=\frac{4a}{1+4a^2}. \ee
When $\kappa=0$, it gives $a=0$. When $\kappa\ne 0$, we have
\bea a=\frac{\pm1-\sqrt{1+\kappa^2}}{2\kappa}. \label{AdS3-solution-1}\eea
We also notice that when we choose the plus sign in the above equation, the limit of $\kappa\to 0$ gives $a\to 0$. We now discuss the BPS condition in the special case when $f=\kappa y$. Now $\G_{M5}$ becomes
\begin{eqnarray} \Gamma_{M5}=\frac{1}{\sqrt{1+\kappa^2}}\big(\kappa\Gamma_{\ul{01245\sharp}}+\Gamma_{\ul{01345\sharp}}-2a(\kappa \Gamma_{\ul{012}}+\Gamma_{\ul{013}})\big)-2a\Gamma_{\ul{45\sharp}}. \end{eqnarray}
After some computation using Eq.~(\ref{ks2}) and the projection conditions Eq.~(\ref{projector2}), we find that $\G_{M5}\eta=\eta$
is equivalent to the projection conditions \be \Gamma_{\ul{2}}\eta_+=\mp\eta_+, \quad \G_{\ul{2}}\eta_-=\mp\eta_-.\ee
The signs on the right side hand of the above two equations follow the choice of the sign in  Eq.~\eqref{AdS3-solution-1}.
Since these projection conditions are compatible with the projection conditions in Eq.~(\ref{projector2}), this M5-brane solution is half-BPS.

In \cite{Yamaguchi}, the M5-brane with worldvolume $AdS_3\times M_3$ embedded in $AdS_4\times M_7$ has been shown to be half-BPS provided that $M_7$
is a weak $G_2$ manifold and $M_3$ is an associate submanifold. Consider the three-form
\be \Phi=\frac{1}{32}\bigg(d\psi+\sum_{i=1}^3\cos\theta_i d\phi_i\bigg)\wedge \sum_{i=1}^3 \bigg(d\theta_i\wedge \sin\theta_id\phi_i\bigg)\ee
in $Q^{1, 1, 1}$, one can easily show that  \be d\Phi=-4\ast\Phi\ee
and \be \Phi|_{M_3}=d\,vol_{M_3}.\ee
This shows explicitly that $Q^{1, 1, 1}$ is a weak $G_2$ manifold, as Sasaki-Einstein manifolds are special cases of $G_2$ manifolds, and the $M_3$ used here is in fact an associate submanifold.
So our results are consistent with the ones in \cite{Yamaguchi}. We also notice that similar BPS M5-branes with worldvolume $AdS_3\times S^3$
in $AdS_4\times S^7$  were studied in  \cite{Lunin:2007ab, Chen:2007tt}.

\section{Discussions}
In this work, we studied some solutions of  the complicated M5-brane equations of motion in the M-theory background $AdS_4\times Q^{1, 1, 1}$.
For the two M5-brane solutions whose worldvolme has an $AdS_2$ factor, we found that both of them are  non-BPS by studying the projection conditions.
Our experiences indicate that there seems to be no BPS M5-branes with such an $AdS_2$ factor. It would be interesting to establish such general
no-go results for $AdS_4\times Q^{1, 1, 1}$ and  more general backgrounds with 8 supercharges. It is also interesting to study such M5-branes with an $AdS_2$ factor
in $AdS_4\times Y^7$, with $Y^7$ a $3$-Sasakian manifold (an M5-brane with an $AdS_3$ factor in $AdS_4\times N(1, 1)$ was studied in \cite{Fujita:2010pj}).

The M5-branes with an $AdS_3$ factor \cite{Ahn:1999ec, Yamaguchi, Lunin:2007ab, Chen:2007tt, Fujita:2010pj}
are believed to dual to some domain walls in the field theory.
It will be interesting to give more concrete description of these domain walls since now we know much more about the dual superconformal field theory. The BPS nature of these M5-branes would allow us to establish the detailed correspondence between the computations in the bulk theory and in the boundary field theory.

\vspace{1.5cm}
\noindent {\large{\bf Acknowledgments}} \\~\\
Z.~L. would like to thank Professor Chuan-Jie Zhu for his generous support, guidance and encouragement.
He would also like to thank the ICTP for financial support for participating in {\it `Spring School on Superstring Theory and Related Topics'}.
J.~W. would like to thank Meng-Qi Zhu for collaboration on related topics and ICTS-USTC for  warm hospitality during a recent visit.
This work was supported in part by NSFC Grants No.~11275010(B.~C.), No.~11335012(B.~C.), No.~11325522(B.~C.), No.~11105154(D.~L. and J.~W.), No.~11222549(D.~L. and J.~W.) and No.~11135006(Z.~L.).
 J.~W. gratefully acknowledges the support of K.~C.~Wong
Education Foundation and Youth Innovation Promotion Association of CAS.

\section*{Appendix: Connection coefficients}

The spin connections with respect to the vielbeins (\ref{vielbein}) are
\bea
 \omega^{\ul{0}\ul{1}}=\frac{2}{R}\frac{\tanh\rho}{\cosh u}e^{\ul{0}},& &
  \omega^{\ul{0}\ul{2}}=\frac{2}{R}\tanh u e^{\ul{0}},\\
  \omega^{\ul{1}\ul{2}}=\frac{2}{R}\tanh u e^{\ul{1}},& &
  \omega^{\ul{2}\ul{3}}=-\frac{2}{R} \coth u e^{\ul{3}},\\
  \omega^{\ul{4}\ul{5}}=\frac{1}{R}(-2\sqrt{2}\cot\theta_1 e^{\ul{5}}+e^{\ul{\sharp}}),& &
  \omega^{\ul{6}\ul{7}}=\frac{1}{R}(-2\sqrt{2}\cot\theta_2 e^{\ul{7}}+e^{\ul{\sharp}}),\\
  \omega^{\ul{8}\ul{9}}=\frac{1}{R}(-2\sqrt{2}\cot\theta_3 e^{\ul{9}}+e^{\ul{\sharp}}),& &
 \omega^{\ul{4}\ul{\sharp}}=\frac{1}{R}e^{\ul{5}},\\ \omega^{\ul{5}\ul{\sharp}}=-\frac{1}{R}e^{\ul{4}}, & &
  \omega^{\ul{6}\ul{\sharp}}=\frac{1}{R}e^{\ul{7}},\\ \omega^{\ul{7}\ul{\sharp}}=-\frac{1}{R}e^{\ul{6}},& &
   \omega^{\ul{8}\ul{\sharp}}=\frac{1}{R}e^{\ul{9}},\hs{3ex} \omega^{\ul{9}\ul{\sharp}}=-\frac{1}{R}e^{\ul{8}}.
 \eea

 The Levi-Civita connection coefficients of the induced metric (\ref{induced}) are
\bea {\G}^t_{~t\rho}&=&\tanh\rho,\hs{3ex}
     \G^\rho_{~tt}=\sinh\rho\cosh\rho,\\
     \G^{\theta_1}_{~\phi_1\phi_1}&=&-\frac12\sin\theta_1\cos\theta_1,\\
     \G^{\theta_1}_{~\phi_1\phi_2}&=&\frac14 \sin\theta_1\cos\theta_2,\\
     \G^{\theta_2}_{~\phi_2\phi_2}&=&-\frac12\sin\theta_2\cos\theta_2,\\
     \G^{\theta_2}_{~\phi_1\phi_2}&=&\frac14 \sin\theta_2\cos\theta_1,\\
          \G^{\phi_1}_{~\theta_1\phi_1}&=&\frac{\sin 2\theta_1(\cos 2\theta_2 -7)  }
 {8 (\cos 2\theta_1 +\cos 2\theta_2-2)},\\
     \G^{\phi_1}_{~\theta_2\phi_1}&=&-\frac{\cos ^2\theta_1  \sin 2\theta_2 }
 {4 (\cos 2\theta_1 +\cos 2\theta_2-2)},\\
 \G^{\phi_1}_{~\theta_1\phi_2}&=& -\frac{\sin\theta_1\cos\theta_2  (\cos 2\theta_2 -3)  }
 {4 (\cos 2\theta_1+\cos 2\theta_2 -2)},\\
 \G^{\phi_1}_{~\theta_2\phi_2}&=& \frac{\cos\theta_1 \sin\theta_2 (\cos 2\theta_2 +5)  }
 {4 (\cos 2\theta_1 +\cos 2\theta_2 -2)},\\
  \G^{\phi_2}_{~\theta_2\phi_2}&=&\frac{\sin 2\theta_2(\cos 2\theta_1 -7)  }
 {8 (\cos 2\theta_1 +\cos 2\theta_2-2)},\\
     \G^{\phi_2}_{~\theta_1\phi_2}&=&-\frac{\cos ^2\theta_2  \sin 2\theta_1 }
 {4 (\cos 2\theta_1 +\cos 2\theta_2-2)},\\
 \G^{\phi_2}_{~\theta_2\phi_1}&=& -\frac{\sin\theta_2\cos\theta_1  (\cos 2\theta_1 -3)  }
 {4 (\cos 2\theta_1+\cos 2\theta_2 -2)},\\
 \G^{\phi_2}_{~\theta_1\phi_1}&=& \frac{\cos\theta_2 \sin\theta_1 (\cos 2\theta_1 +5)  }
 {4 (\cos 2\theta_1 +\cos 2\theta_2 -2)}.
       \eea

The Levi-Civita connection coefficients of the induced metric (\ref{induced2}) are
\bea
\G^{t}_{~\rho t}=\tanh\rho,& &
\G^{\rho}_{~tt}=\cosh\rho\sinh\rho,\\
\G^{\theta_1}_{~\phi_1\phi_1}=-\frac12\sin\theta_1\cos\theta_1,& &
\G^{\theta_1}_{~\phi_1\psi}=\frac14\sin\theta_1,\\
\G^{\phi_1}_{~\phi_1\theta_1}=\frac34\cot\theta_1,& &
\G^{\phi_1}_{~\theta_1\psi}=-\frac1{4\sin\theta_1},\\
\G^{\psi}_{~\phi_1\theta_1}=-\frac34\cot\theta_1\cos\theta_1-\frac12\sin\theta_1,&&
\G^{\psi}_{~\theta_1\psi}=\frac14\cot\theta_1.\eea

The nonzero components of the Christoffel symbol for the metric (\ref{induced3}) are
\bea \G^{\theta_1}_{~\phi_1\phi_1}=-\frac12\sin\theta_1\cos\theta_1,& &
\G^{\theta_1}_{~\phi_1\phi_2}=\frac14\sin\theta_1\cos\theta_2,\\
\G^{\theta_1}_{~\phi_1\psi}=\frac14\sin\theta_1,& &
\G^{\phi_1}_{~\theta_1\phi_1}=\frac34\cot\theta_1,\\
\G^{\phi_1}_{~\theta_1\phi_2}=-\frac14\csc\theta_1\cos\theta_2,& &
\G^{\phi_1}_{~\theta_1\psi}=-\frac14\csc\theta_1,\\
\G^{\theta_2}_{~\phi_1\phi_2}=\frac14\cos\theta_1\sin\theta_2,& &
\G^{\theta_2}_{~\phi_2\phi_2}=-\frac12\sin\theta_2\cos\theta_2,\\
\G^{\theta_2}_{~\phi_2\psi}=\frac14\sin\theta_2,& &
\G^{\phi_2}_{~\phi_1\theta_2}=-\frac14\cos\theta_1\sin\theta_2, \\
\G^{\phi_2}_{~\phi_2\theta_2}=\frac34\cot\theta_2,& &
\G^{\phi_2}_{~\theta_2\psi}=-\frac14\csc\theta_2,\\
\G^\psi_{~\theta_1\phi_1}=-\frac18\csc\theta_1(\cos2\theta_1+5),& &
\G^\psi_{~\theta_1\phi_2}=\frac14\cot\theta_1\cos\theta_2,\\
\G^\psi_{~\psi\theta_1}=\frac14\cot\theta_1,& &
\G^\psi_{~\phi_1\theta_2}=\frac14\cos\theta_1\cot\theta_2,\\
\G^\psi_{~\theta_2\phi_2}=-\frac18\csc\theta_2(\cos2\theta_2+5),& &
\G^\psi_{~\psi\theta_1}=\frac14\cot\theta_2.
\eea

 The Christoffel symbols of the reduced metric (\ref{induced4}) are
 \bea \G^{t}_{~ty}=-\frac1y, & & \G^{x_1}_{~x_1y}=-\frac1y, \\ \G^y_{~tt}=-\G^{y}_{~x_1x_1}=-\frac{1}{y(1+f^{\prime 2})},& &
 \G^{y}_{~yy}=\frac{f^{\prime}f^{\prime\prime}}{1+f^{\prime2}}-\frac1y, \\
\G^{\theta_1}_{~\phi_1\phi_1}=-\frac12\sin\theta_1\cos\theta_1,& &
\G^{\theta_1}_{~\phi_1\psi}=\frac14\sin\theta_1,\\
\G^{\phi_1}_{~\phi_1\theta_1}=\frac34\cot\theta_1,& &
\G^{\phi_1}_{~\theta_1\psi}=-\frac1{4\sin\theta_1},\\
\G^{\psi}_{~\phi_1\theta_1}=-\frac34\cot\theta_1\cos\theta_1-\frac12\sin\theta_1,& &
\G^{\psi}_{~\theta_1\psi}=\frac14\cot\theta_1.\eea


\end{document}